# Image Domain Dual Material Decomposition for Dual-Energy CT using Butterfly Network

Wenkun Zhang[1], Hanming Zhang[1], Linyuan Wang[1], Xiaohui Wang[2], Ailong Cai[1], Lei Li[1, a)],

Tianye Niu[3, 4, a)], Bin Yan[1]

[1]*National Digital Switching System Engineering and Technological Research Center, Zhengzhou 450002,*

*China*

[2]*People's Liberation Army of China Number 153 Central Hospital, Zhengzhou 4500002, China*

[3]*Sir Run Run Shaw Hospital, Zhejiang University School of Medicine; Institute of Translational Medicine,*

*Zhejiang University, Hangzhou, Zhejiang, 310009, China*

[4]*Key Laboratory of Biomedical Engineering of Ministry of Education, Zhejiang University, Hangzhou,*

*Zhejiang,310009, China*

a) Author to whom correspondence should be addressed. Electronic mails:

leehotline@aliyun.com (Lei Li), and tyniu@zju.edu.cn (Tianye Niu).

**Purpose:** Dual-energy CT (DECT) has been increasingly used in imaging applications because of its capability for material differentiation. However, material decomposition suffers from magnified noise from two CT images of independent scans, leading to severe degradation of image quality. Existing algorithms achieve suboptimal decomposition performance since they fail to accurately depict the mapping relationship between DECT and the basis material images. Convolutional neural network (CNN) exhibits great promise in the modeling of data coupling and has recently become an important technique in medical imaging application. Inspired by its impressive potential, we developed a new Butterfly network to perform the image domain dual material decomposition due to its strong approximation ability to the mapping functions in DECT.

**Methods:** The Butterfly network is derived from the image domain DECT decomposition model by exploring the geometric relationship between mapping functions of data model and network components. The network is designed as the double-entry double-out crossover architecture based on the decomposition formulation. It enters a pair of dual-energy images as inputs and defines the ground true decomposed images as each label. The crossover architecture, which plays an important role in material decomposition, is designed to implement the information exchange between the two material generation pathways in the network. The evaluation consists of network visualization and basis material decomposition. We disassemble the components of network to evaluate their sensitivity to different basis materials. The proposed network is further applied on clinical DECT images to evaluate its performance in clinical applications.

**Results:** Network components exhibit different sensitivity to basis materials in the visualization. By analyzing their sensitivity to different basis materials, we determined the roles of network components in material decomposition. This visualization evaluation reveals what the network can learn and verifies the rationality of network design. Besides, the qualitative and quantitative evaluations in material decomposition of patient data indicate that the proposed network outperforms its counterpart. The proposed network reduces the standard deviation of the noise in decomposed images by over 90% and 80% compared with that using the direct matrix inversion and conventional iterative method, respectively.

**Conclusions:** We develop a model-based Butterfly network to perform image domain material decomposition for DECT. The results validate its capability of recognizing different basis materials from DECT images. The proposed approach also leads to higher decomposition quality in noise suppression on clinical datasets as compared with those using conventional schemes.

# I. INTRODUCTION

Dual-energy CT (DECT) has been increasingly used in advanced imaging applications, including medical imaging[1,2], security inspection[3,4], and nondestructive testing[5]. Compared with conventional single-spectrum CT technology, DECT requires the acquisition of two datasets at two distinct x-ray energy spectra. DECT can characterize different materials on the basis of their elemental composition[6] and exceed the physical limits of conventional single-energy CT with its advantage of material differentiation[7].

DECT reconstruction approximates the energy-dependent attenuation in each voxel using a linear combination of photoelectric absorption and Compton scattering for most materials[8]. In practice, it is more convenient to parameterize the energy-dependent attenuation as a linear combination of linear attenuation coefficients of two basis materials[9]. Existing approaches for DECT can be divided into three types[10]: direct reconstruction, projection domain based, and image domain based methods. Direct reconstruction approaches reconstruct basis images directly from data collected at two x-ray energy spectra[11,12]. This type of methods can depict the data model of DECT and incorporate regularization constrains, leading a great degree of flexibility for accommodating different CT scanners. However, this approach requires high computational complexity, and the decomposition results are sensitive to parameters selection in different applications[13]. Projection domain based methods first convert the measured projections into independent sinograms of basis materials using, e.g., polynomial fitting[8], and then reconstruct the DECT images using standard reconstruction algorithms[14]. This type of methods can appropriately compensate for non-linear effect in material decomposition of projection domain, and correct for the beam-hardening artifacts in the decomposition process[15]. Projection domain method highly depends on spatial geometric consistency of measured projections, which is a challenge for the imaging systems such as the dual-source and fast voltage-switching DECT scanners[16]. Unlike the projection domain methods, image domain-based decomposition first reconstructs dual-energy images individually or jointly from two measured datasets, subsequently forming the basis image by a linear combination of the reconstructed images[17]. It is more convenient to directly apply on the DECT images acquired from commercial CT scanner compared with the other two methods[18,19]. One generic problem of current DECT is that the noise magnification of the decomposed material images due to the noise correlation in the high- and low-energy CT images, resulting in the sensitive of decomposition to the noise.[20,21]. Standard digital image processing techniques (e.g., low-pass filtering) are first used as an auxiliary step to suppress noises but have limited efficacy[22,23]. Advanced algorithms utilize redundant structural or statistical information of the CT or decomposed images for noise suppression, with improvements on noise reduction in DECT[24,25]. For example, ref. [26] achieved superior performance for noise suppression by developing an iterative image-domain decomposition method using the full variance-covariance matrix of the decomposed images. However, these existing algorithms do not fully depict the mapping relationship between DECT images and basis material images. In this paper, we develop a novel approach based on convolutional neural network (CNN) to improve the quality of material decomposition due to its strong approximation ability to mapping functions.

Deep learning techniques receive considerable attention in the field of x-ray imaging. Review paper [27] first summarized the relationship between machine learning and medical imaging. It pointed out that the inverse function for various tomographic modalities are representable using neural network. The initial applications of machine learning in image reconstruction mainly focus on the noise suppression of low-dose CT[28-30]. Researchers mapped low-dose CT images towards its corresponding normal-dose counterparts in spatial[31] or transformation domain[32,33] via CNN, which showed a substantial improvement over conventional noise suppression approaches. However, CNN was mainly used for post-processing on the CT images in these methods and the procedure of image reconstruction was not investigated to improve the network design for specific CT problems. Recently, researchers start to investigate the link between conventional algorithms and deep learning networks[34-36]. They provided a new perspective on the design of effective networks by applying CNN as an operator in the conventional approaches. Ref. [37] recently further bridged the gap between deep neural network design and numerical differential equations. Many effective networks can be interpreted as different numerical discretization of differential equations. These investigations indicate an intrinsic connection between conventional approaches and neural network. The connection is crucial for the interpretation of capability of the networks in realizing the alternative functionality for conventional approaches and guidance in developing specific networks for practical imaging problems.

Most of aforementioned works for x-ray imaging merely consider CNN as a tool to realize some functions without design philosophy under a guideline scheme. They may depart from CT data model which is the mathematical and physical basis of CT imaging. Designing a network based on data model or embedding a tight connection into data model is significant to fully optimize the performance of networks. In this work, we investigate the relationship between image domain decomposition model and neural network. A new model-based CNN is developed to perform material decomposition in image domain for DECT. Unlike existing methods in x-ray imaging, the proposed network is not a direct application of existing network. Instead, the procedure by which the formulation of the material decomposition is transformed into a new network architecture. Inspired by the design philosophy of channel coding model[38], a crossover structure is proposed in the new network to perform information exchange between DECT images. By disassembling the components of the network, we verify the rationality of the network design and interpret the components of the trained network and their roles in material decomposition. Finally, the effectiveness of our method is validated using the results of material decomposition on clinical DECT data. The novel network can learn comprehensive and abstract knowledge of basis material from big training data. This design enables the network to learn and recognize different basis material from DECT images, and exhibit promising performance in material decomposition.

## II. METHOD

### II. A. Image Domain Material Decomposition

In DECT, the linear attenuation coefficient of each pixel in the CT image is approximated by the linear combination of the pixel values in the images of basis materials based on the theory of image domain decomposition[39]. In this paper, the model of the material composition can be written as:

$$\begin{pmatrix} \mu_{\mathrm{L}} \\ \mu_{\mathrm{H}} \end{pmatrix} = \begin{pmatrix} \mu_{1\mathrm{L}} & \mu_{2\mathrm{L}} \\ \mu_{1\mathrm{H}} & \mu_{2\mathrm{H}} \end{pmatrix} \begin{pmatrix} x_1 \\ x_2 \end{pmatrix}, \tag{1}$$

where the subscript $\mathrm{L}$ and $\mathrm{H}$ represents the low- and high-energy spectrum, respectively,

and subscripts 1 and 2 indicate two basis materials. $\mu_{ij}$ is the linear attenuation coefficient of material $i \in \{1,2\}$ at the energy spectrum $j \in \{L,H\}$. $x_1, x_2$ are the normalized volume fractions of the basis materials at the same position of two basis material images. $\mu_L, \mu_H$ is the pixel pair of the DECT images. Let $A$ represent the material composition matrix with a dimension of $2N \times 2N$, where $N$ is the total number of pixels in one complete image. The composition matrix $A$ is denoted as following:

$$A = \begin{pmatrix} \mu_{1L}I & \mu_{2L}I \\ \mu_{1H}I & \mu_{2H}I \end{pmatrix}. \tag{2}$$

where $I$ is an identity matrix with the dimension $N \times N$. We can then obtain basis material images from dual-energy images via matrix inversion. Such direct material decomposition is written as follows:

$$\vec{x} = A^{-1}\vec{\mu}. \tag{3}$$

In Eq. (3), $\vec{\mu}$ is a $2N$ vector composed of concatenated low- and high-energy images $\vec{\mu}_L, \vec{\mu}_H$. $\vec{x}$ is a $2N$ vector consisting of the concatenated decomposed images of basis materials $\vec{x}_1, \vec{x}_2$. Inverse operation $A^{-1}$ represents decomposition matrix, which is calculated analytically as following:

$$A^{-1} = \begin{pmatrix} A_{11}I & A_{12}I \\ A_{21}I & A_{22}I \end{pmatrix} = \frac{1}{\mu_{1L}\mu_{2H} - \mu_{2L}\mu_{1H}} \begin{pmatrix} \mu_{2H}I & -\mu_{2L}I \\ -\mu_{1H}I & \mu_{1L}I \end{pmatrix} \tag{4}$$

with the components $A_{11} = \mu_{2H} / (\mu_{1L}\mu_{2H} - \mu_{2L}\mu_{1H})$, $A_{12} = -\mu_{2L} / (\mu_{1L}\mu_{2H} - \mu_{2L}\mu_{1H})$, $A_{21} = -\mu_{1H} / (\mu_{1L}\mu_{2H} - \mu_{2L}\mu_{1H})$, and $A_{22} = \mu_{1L} / (\mu_{1L}\mu_{2H} - \mu_{2L}\mu_{1H})$.

The procedure of material decomposition is formulated as an ill-posed problem of matrix inversion. Regularization-based models including $\ell_0$-, $\ell_1$-, $\ell_\infty$- constrained minimization are introduced to solve this problem using iterative approaches. These optimization problems can be solved using deep neural network with improvements, in both accuracy and complexity, over conventional algorithms[40-42].

We substitute the inverse matrix $A^{-1}$ in Eq. (3) by a mapping function $D(\Theta)$, where $\Theta$ represents the parameters of material decomposition. The decomposition process is interpreted as a procedure which transforms DECT images into basis materials via mapping function $D(\Theta)$. Eq. (3) can then be rewritten as following:

$$\vec{x} = D(\Theta)\vec{\mu}. \tag{5}$$

To solve this problem, in this paper, CNN are used to approximate the mapping function $D(\Theta)$ for material decomposition because of its promising performance in the modeling of data coupling.

## II. B. Network Design Based on Decomposition Model

Material decomposition is achieved by exploring the difference in attenuation coefficients of DECT images. The decomposition of different basis material based on Eq. (5) can be rewritten as following:

$$\begin{pmatrix} \vec{x}_1 \\ \vec{x}_2 \end{pmatrix} = \begin{pmatrix} D_{1L}(\Theta_{1L}) & D_{1H}(\Theta_{1H}) \\ D_{2L}(\Theta_{2L}) & D_{2H}(\Theta_{2H}) \end{pmatrix} \begin{pmatrix} \vec{\mu}_L \\ \vec{\mu}_H \end{pmatrix}, \tag{6}$$

where $\Theta_{ij}$ represents the parameters of mapping function $D_{ij}$. Based on the number of input and output variables in Eq. (6), the proposed CNN is designed as a double-entry double-out architecture. The architecture of CNN to decompose basis material M1 can then be expressed as in Figure 1(a), where "$\rightarrow$" represents different network components corresponding to $D_{1L}(\Theta_{1L})$ and $D_{1H}(\Theta_{1H})$, respectively. A and B represent the double-entry data of DECT images. M1 and M2 denote the double-output images of basis materials. Similarly, the network architecture for the second basis material decomposition is denoted as Figure 1(b), where two network components are corresponding to $D_{2L}\left(\Theta_{2L}\right)$ and $D_{2H}\left(\Theta_{2H}\right)$. Based on the determined double-entry double-out architecture, the material decomposition model in image domain of DECT can be expressed as in Figure 1(c) by combining two decomposition network architectures of basis materials M1 and M2. A crossover architecture between the two material generation pathways are generated to realize reasonable shunt of material information from DECT images.

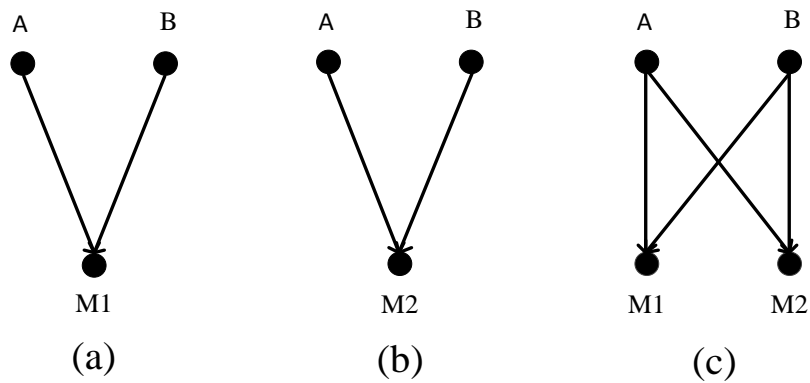

Figure 1 Schematic diagram of network structure. (a) and (b) represent the network architecture of decomposition for material M1 and M2, respectively. (c) Integrated architecture of the proposed CNN for image domain material decomposition in DECT

Figure 2 shows the detailed composition of the new network. Residual blocks[43] are introduced by adding short connections to improve the efficiency of training. The network considers a pair of low- and high-energy images as the double inputs and defines each label as the ground truth of bone and tissue image. The input and output are represented by blue rectangles in the diagram. The network consists of 12 convolutions (green rectangle) followed by batch normalization (BN) (yellow rectangle) and rectified linear unit (ReLU) (purple rectangle), except for the final convolution layer. The brown ellipse of the diagram indicates the combination of addition and ReLU. The whole network architecture is divided into three stages. The first stage involves the feature extraction that transforms the input images into a set of feature maps. In the second stage, SwapA1, Swap A2, Swap B1, and Swap B2 comprise the crossover network. ResA1, ResA2, ResB1, and ResB2 consist of two residual blocks. The final stage recombines the feature maps to whole decomposition results for different basis materials using two convolutions. The network does not have pooling layer so that no down- and up-sampling operations are performed during the process of material decomposition. The new network is named as Butterfly network (Butterfly-Net) according to its shape of double-entry and double-out crossover architecture.

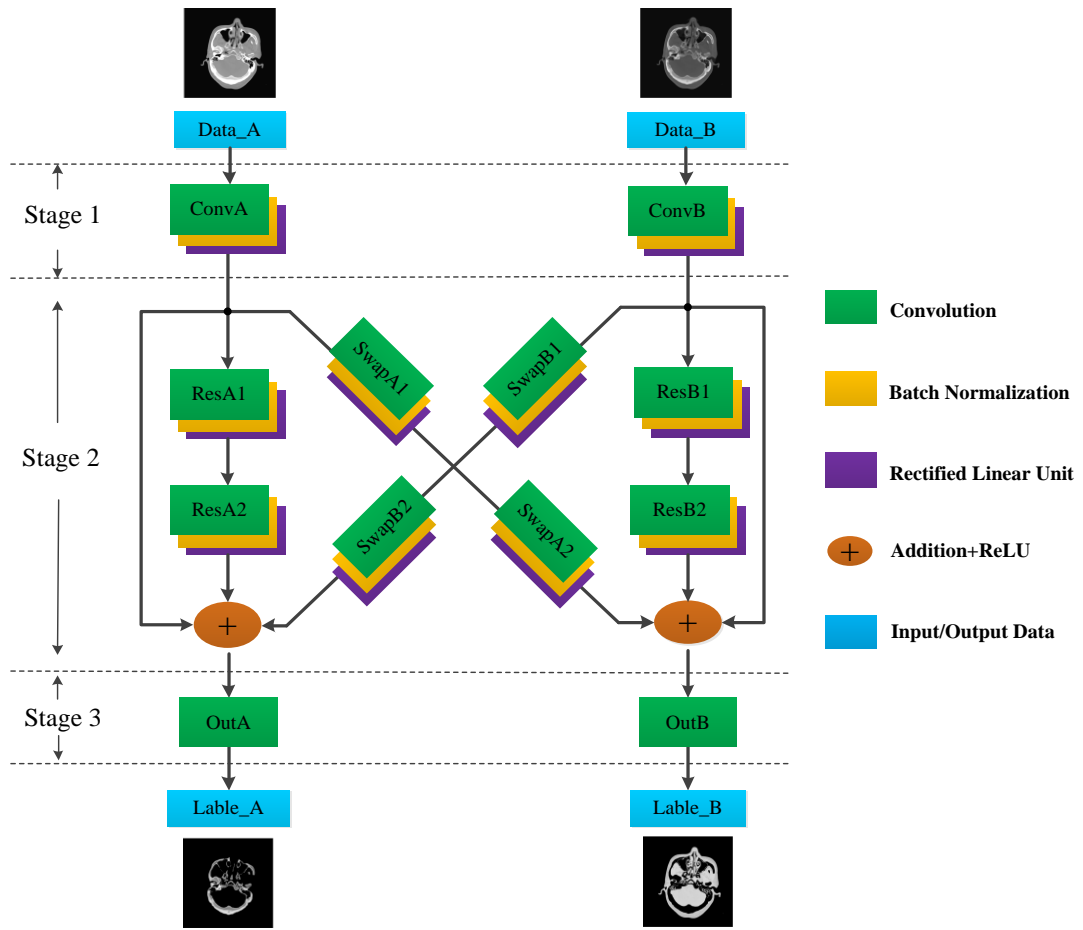

Figure 2 Butterfly-Net architecture. The input and output are indicated by blue rectangles. The green, yellow, purple rectangles represent convolution, BN and ReLU, respectively. The brown ellipse indicates the combined operations of addition and ReLU.

The mathematical representation of the new network is further investigated to analyze the relationship between network and decomposition model. Let $h_{CA}$ and $h_{CB}$ represent the transfer functions of ConvA and ConvB, respectively. The joint ResA1 and ResA2 is indicated by transfer function $h_{RA}$, and the joint ResB1 and ResB2 corresponds to $h_{RB}$. For the crossover network, $h_{SA}$ is the transfer function of joint SwapA1 and Swap A2, and $h_{SB}$ is the transfer function of joint SwapB1 and SwapB2. The transfer function of OutA and OutB are represented as $h_{OA}$ and $h_{OB}$, respectively. Based on the above notations, the mapping formulas of basis materials in the Butterfly-Net can be written as

$$\begin{aligned}\vec{x}_1 &= h_{OA}\left\{\left(1+h_{RA}\right)h_{CA}\left(\vec{\mu}_L\right)+h_{SB}\left[h_{CB}\left(\vec{\mu}_H\right)\right]\right\}, \\ \vec{x}_2 &= h_{OB}\left\{\left(1+h_{RB}\right)h_{CB}\left(\vec{\mu}_L\right)+h_{SA}\left[h_{CA}\left(\vec{\mu}_H\right)\right]\right\},\end{aligned} \tag{7}$$

The transfer functions of Eq. (7) are non-linear since most convolutions in the network are followed by ReLU, which is an activation function of non-linear filtration. Under the assumption that the transfer function of the network is the combination of convolutions, Eq. (7) can be rewritten as following:

$$\begin{aligned}\vec{x}_1 &= h_{1L}\left(\vec{\mu}_L\right)+h_{1H}\left(\vec{\mu}_H\right), \\ \vec{x}_2 &= h_{2L}\left(\vec{\mu}_L\right)+h_{2H}\left(\vec{\mu}_H\right),\end{aligned} \tag{8}$$

where $h_{ij}$ represents the cascade of transfer functions. In this way, we can find that the representation of the proposed network in terms of mathematical formula is consistent with the image domain decomposition model of Eq. (6). The network can be considered as the complete expression of the data model under the assumption that the transfer functions exclude ReLUs.

## III. EVALUATION

### III. A. Network Parameters

We set the parameters of the proposed network according to ref. [44], which gives the detailed comparison of different parameter settings. In the first stage, we apply 64 convolutional kernels with a $7 \times 7$ kernel size on the input dual-energy images. In the second stage, all convolutions are $64 \times 3 \times 3 \times 64$ in size. At the end of this stage, the feature data from different material generation pathways is merged into the integrated 64 channels of feature map. In the final stage, the size of two output convolutions is set as $64 \times 5 \times 5 \times 1$. The standard stochastic gradient descent with momentum is applied in the optimization. The initial learning rate, the momentum, and the weight decay are set as $10^{-6}$, 0.95, and 0.0005, respectively.

## III. B. Training and Testing Data

A total of 130 pairs of bone and tissue images derived from patient CT images are chosen as the labels of Butterfly-Net and the size of the image with normalized densities of the basis materials is $512 \times 512$. We first simulate dual-energy spectra with an energy sampling interval of 1 KeV at tube voltages of 80 and 140 KVp as shown in Figure 3(a). We further generate dual-energy projections for each slice of head images, which is the composition of a pair of bone and tissue images. The attenuation coefficients of bone and tissue are shown in Figure 3(b). Sidden's ray-tracing algorithm is applied to simulate the fan-beam geometry. The source-to-object and source-to-detector distances are 1000 and 1500 mm to mimic the geometry of the clinical on-board imager installed on Varian Trilogy treatment machine (www.varian.com). The dual-energy projections are uniformly sampled in 720 views over 360-deg rotation. Projection samples in each view are collected using a linear detector consisting of 1024 bins with each pixel of 0.388 mm. 130 pairs of high- and low-energy images are reconstructed by performing filtered backprojection (FBP) algorithm on the simulated dual-energy projections. All reconstructed dual-energy images are composed of 512×512 square pixels with physical size of 0.259×0.259 mm$^2$ for each pixel. These images are taken as the input data of the Butterfly-Net for training. We partition DECT and basis tissue images into the overlapping 128×128 image patches with a sliding window of 48 to efficiently boost the number of samples. The dataset finally consists of 10530 pairs of image patches, 9720 and 810 of which are picked up as the training and validation set.

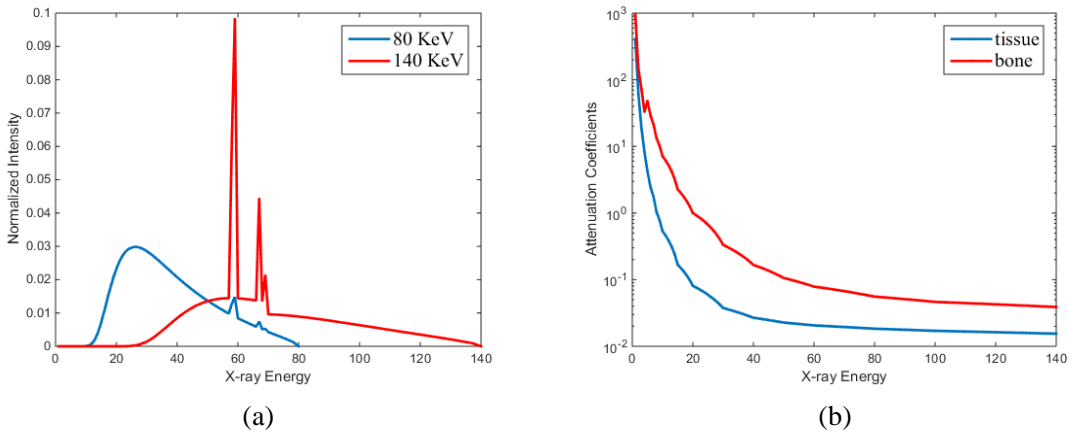

(a)                                                                (b)

Figure 3 Dual-energy spectrum and attenuation coefficients of bone and tissue. (a) The blue and red lines denote the spectrum of 80 and 140 KVp with energy incremental interval of 1 KeV, respectively. (b) The blue and red lines are the attenuation coefficients of tissue and bone within 140 KVp, respectively.

Two evaluations are performed in this paper. One is the visualization evaluation to show the process of basis material decomposition using the network. The other one is the decomposition study to validate the performance of our method using clinical data. Simulation phantom and patient data are used to perform the network visualization. We firstly construct a digital phantom filled with bone and tissue material. As shown in Figure 4(a), the phantom has three discrete grey levels including the background value of 0. We perform visualization study in the bone and tissue region to evaluate the influence of different material information to the network. Dual-energy projections at a tube voltage of 80 and 140 KVp can be obtained using the Sidden's ray-tracing algorithm. The set of parameter for the simulation are consistent with that of the training datasets. Noise is added to the dual-energy projections, which are generated using a Poisson model as following:

$$N = \text{Poisson}\left(N_0 \exp\left(-p\right)\right), \tag{9}$$

where $N_0$ is the number of incident x-ray photons and $p$ denotes the measured number of photons in the projection. $N_0$ was set as $5 \times 10^5$ according to ref. [45]. The DECT images are reconstructed using FBP algorithm and they are shown in Figure 4(b) and (c).

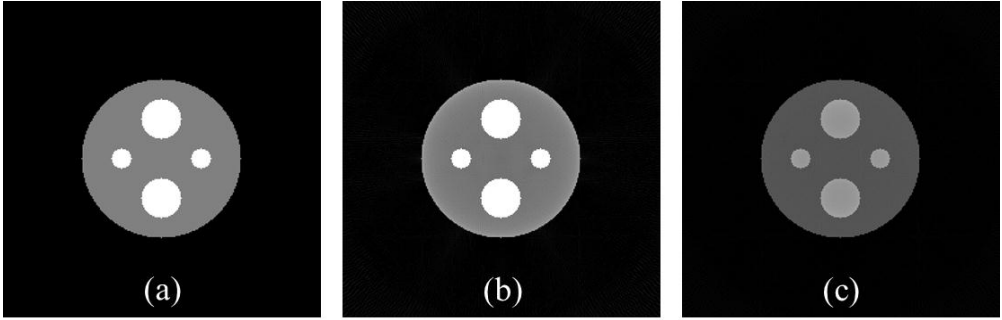

Figure 4 Digital phantom for visualization evaluation. (a) represents the simulation phantom; (b) and (c) are the reconstructed images of the simulation phantom at tube voltage of 80 and 140 KVp with display window of [0 0.6] cm$^{-1}$, respectively.

Patient head data are directly collected from the medical DECT system of SIEMENS at tube voltage of 80 and 140 KVp, which are provided by Radiology Department in a local hospital. The DECT images have a dimension of 512×512. Figure 5(a) and (b) show one pair of the low- and high-energy head image used to perform visualization evaluation. Different from the simulation evaluation, bone and tissue information are merged in CT images which results in the mixed pixels containing both bone and tissue information. Since it is difficult to strictly and accurately separate bone and tissue regions, we appropriately divide two regions based on the rough distribution of bone and tissue materials. Region I (Figure 5(c)) mainly contains bone material and Region II (Figure 5(d)) contains mostly tissue information. We also compare the decomposition results of the proposed network using clinical head data with the direct matrix inversion technique and conventional method in ref. [26]. The mean and standard deviation (SD) of a selected region of interest (ROI) is used to quantify the noise level.

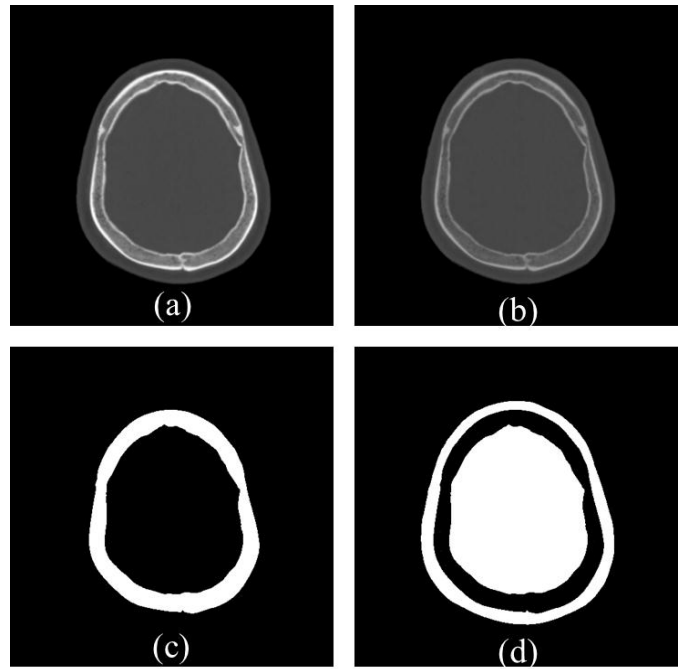

Figure 5 (a) and (b) are the reconstructed images at tube voltage of 80 and 140 KVp with display window of [0 0.05] mm$^{-1}$, respectively; (c) and (d) are the regions I and II for network visualization, respectively.

### III. C. Network Components and Visualization

Sensitivity analysis is applied to evaluate the activation level of neurons in the network. An efficient tool for sensitivity analysis is the heatmap method[46], which measures the small changes in the pixel value with respect to the network output. The heatmap expresses the sensitivity of neurons to the input data[47]. In this paper, we identify the part of the neurons in the network which is sensitive to bone and tissue basis materials. Hence, heatmap is properly adopted for the testing of the sensitivity of network to different basis materials.

To compare the heatmaps of different network components, the Butterfly-Net is assembled into four network components (Figure 6) based on the corresponding relationship of network components and mapping functions. By investigating the sensitivity of network components to different material information, we can analyze the learning process of the network component after training and further interpret the decomposition mechanism of Butterfly-Net. For each diagram in Figure 5, the colorized neurons are composed of the disassembled components to evaluate their sensitivity to basis materials. The $1^{st}$ and $2^{nd}$ components connect the input low-energy image to the labels of bone and tissue images, respectively. The two network components mainly test their sensitivity to different basis materials from a low-energy image. The $3^{rd}$ and $4^{th}$ components link the input high-energy image to the labels of bone and tissue images, respectively. The two network components mainly test their sensitivity to different basis materials from a high-energy image. The $2^{nd}$ and $3^{rd}$ components comprise the crossover architecture, which plays the key role for material decomposition in Butterfly-Net. Their heatmap results show the decomposition process of basis materials from DECT images.

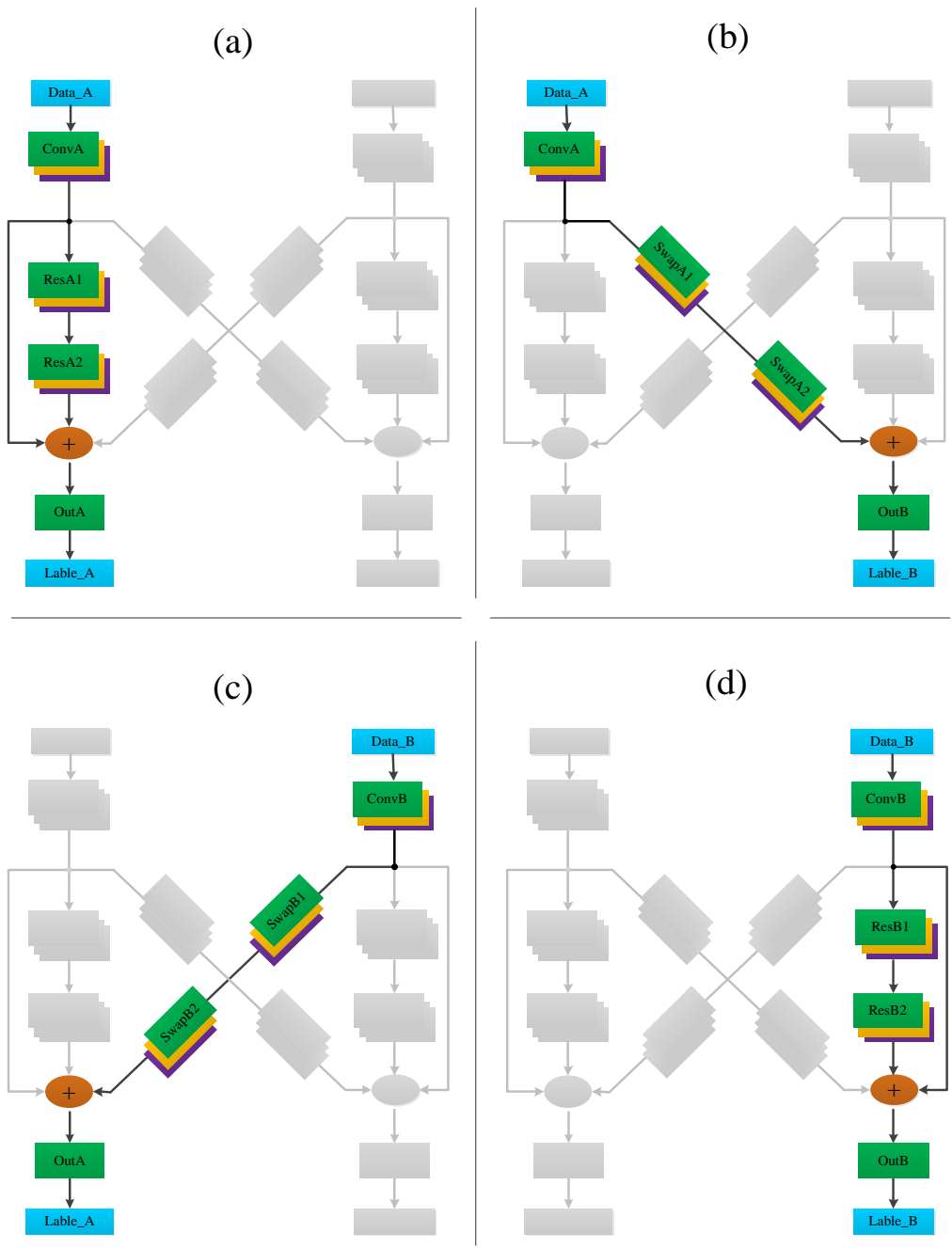

Figure 6 The disassembled components of Butterfly-Net. The colorized convolutions are investigated and the gray convolutions are the silent ones. (a), (b), (c), and (d) represent the 1st, 2nd, 3rd, and 4th network components, respectively.

# IV. RESULTS

## IV. A. Digital phantom study

DECT images of the digital phantom are loaded into the trained Butterfly-Net to perform the visualization study. In Figure 7, the top and bottom rows represent the heatmaps in bone and tissue regions, respectively. The results from left to right labeled from ① to ④ correspond to the $1^{st}$ to $4^{th}$ network components. The heat values of the $1^{st}$ and the $3^{rd}$ components are higher in the bone region than the tissue region. The excitability of their neurons increases when bone information of dual-energy images flows through the two components. Compared with the bottom results of the $1^{st}$ and $3^{rd}$ components, neurons are not activated or are less active when tissue information passes through the two components. This finding indicates that the $1^{st}$ and the $3^{rd}$ components are more sensitive to bone information than to tissue information of the input DECT images. A similar phenomenon is observed in the $2^{nd}$ and the $4^{th}$ components. Activity of the neurons increases when tissue information rather than bone information flows into the two components, which means that the $2^{nd}$ and the $4^{th}$ components are more sensitive to tissue information than to bone information.

Based on the visualization results of the digital phantom, the disassembled components of network exhibit different sensitivities to bone and tissue basis materials. The $1^{st}$ and the $3^{rd}$ components exhibit higher sensitivity to bone information, whereas the $2^{nd}$ and the $4^{th}$ components are more active when tissue information passes through them. Based on the joint effect of four components, the integrated network can recognize different basis materials. The $2^{nd}$ and $3^{rd}$ components comprise the crossover structure of Butterfly-Net, which facilitate the information exchange between two material generation pathways. This finding exactly tallies with the physical mechanism of decomposition model that utilizes the difference of DECT images to decomposed materials.

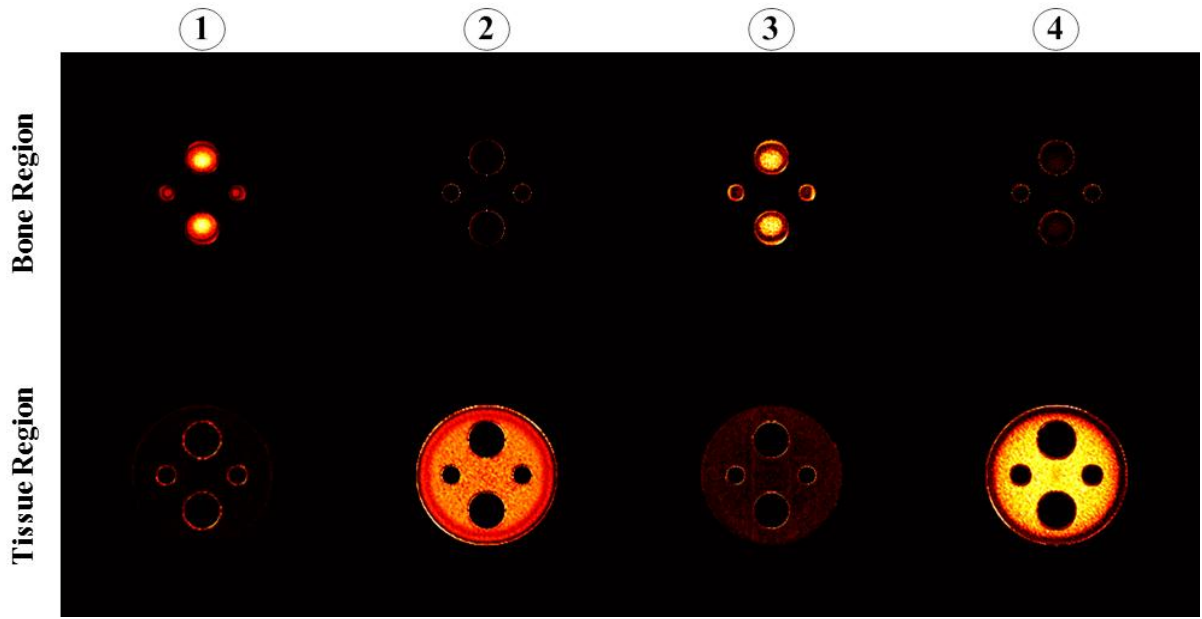

Figure 7 Heatmap of different components for the digital phantom. The top and bottom row represents the heatmaps in bone and tissue regions. The results from left to right correspond to the $1^{st}$ to $4^{th}$ network components.

## IV. B. Head patient study

### IV. B. 1 Visualization Results

Visualization evaluation is also performed on clinical head data. Heatmaps of different network components are shown in Figure 8. The heat values in region I are high for all components, because this region contains both bone and tissue information. Comparing the distribution positions of heat pixels in detail for different network components, the distribution was found to be highly dependent on the output labels of the components. For example, the pixels in heatmaps of the $2^{nd}$ and $4^{th}$ network components, which point to the tissue label, are mostly gathered in the tissue area of region I. By comparing the heatmaps of the components with the same input, e.g., the $1^{st}$ and $2^{nd}$ components or the $3^{rd}$ and $4^{th}$ components, it was found that most of the heat values of the tissue area are higher than those of other areas in region I. Thus, to distinguish tissue information from region I, the neurons that recognize tissue information must become more active than the other neurons, which are sensitive to background information. This phenomenon depends on the experience so much that researchers must be more sensitive to the object material to recognize it from the mixed material. The heatmaps in bone and tissue areas are continually compared for one network component. A phenomenon similar to that observed in the simulation evaluation is also observed. The $1^{st}$ and $3^{rd}$ components exhibit higher sensitivity to the bone area than to the tissue area. The heat values of tissue area in the $2^{nd}$ and $4^{th}$ components, which are simultaneously distributed in both regions I and II, are higher than most of the bone area.

Visualization evaluation of both simulation and patient data enables the determination of roles of different network components in material decomposition. It reveals how the network performs material decomposition and verifies that the trained Butterfly-Net learns and decomposes basis materials from input DECT images. Hence, the philosophy of network design is rational for image domain material decomposition.

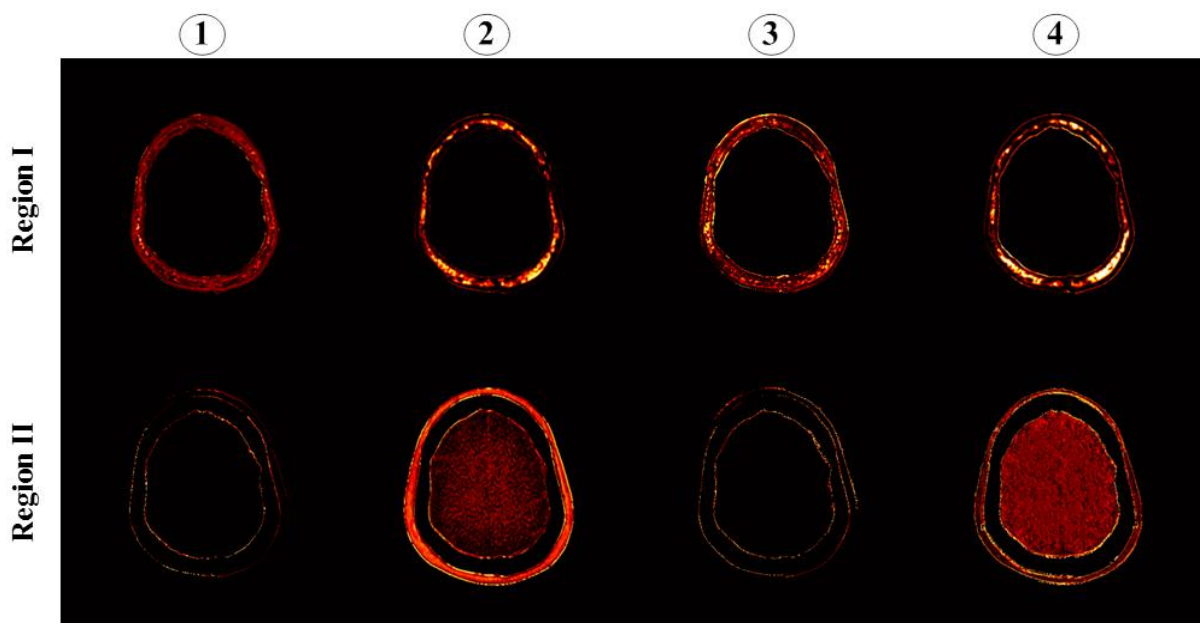

Figure 8 Heatmap of different components for clinical head data. The top and bottom row represents the heatmaps in region I and II. The results from left to right correspond to the 1st to 4th network components.

*IV. B.2 Comparison of Material Decomposition*

Material decomposition is performed on clinical DECT images from 10 patients at a tube voltage of 80 and 140 KVp to compare the performance of different approaches. Three pairs of DECT images are randomly selected in the testing dataset to qualitatively and quantitatively compare decomposition results of different methods. The three groups of head DECT data are not included in the training datasets of Butterfly-Net.

Figures 10, 11, and 12 show the decomposition results of different methods using clinical data. The first column presents the DECT images of the imaging system. The second column presents the results of direct matrix inversion. The result of this method has high noise in the decomposed images. The third column illustrates the decomposed bone and tissue images using the iterative method in ref. [26], in which image noise is suppressed and relatively higher image quality is reached. The fourth column shows the decomposed images using Butterfly-Net. Compared with the former methods, the proposed method reduces the noise on the bone and tissue images to a greater extent.

The mean and standard deviation (SD) are calculated on the decomposed images generated by different algorithms. The ROI for quantitative analysis is represented by red rectangles in different groups. The quantitative results of decomposition images are shown in Table 1. The values of the decomposed images are unitless, because they denote the normalized densities of the basis materials. Our method reaches a higher SNR and reduces the noise SD compared with the other two methods. Compared with the direct matrix inversion, the proposed method reduces the noise SD of decomposed images by 93.3%, 93.1%, and 91.3% in the three groups of head data. Compared with the conventional iterative method, the SD is reduced by 81.3%, 85.8%, and 80.3%. Therefore, the quantitative results also indicate that our network can decompose material with increased quality for DECT.

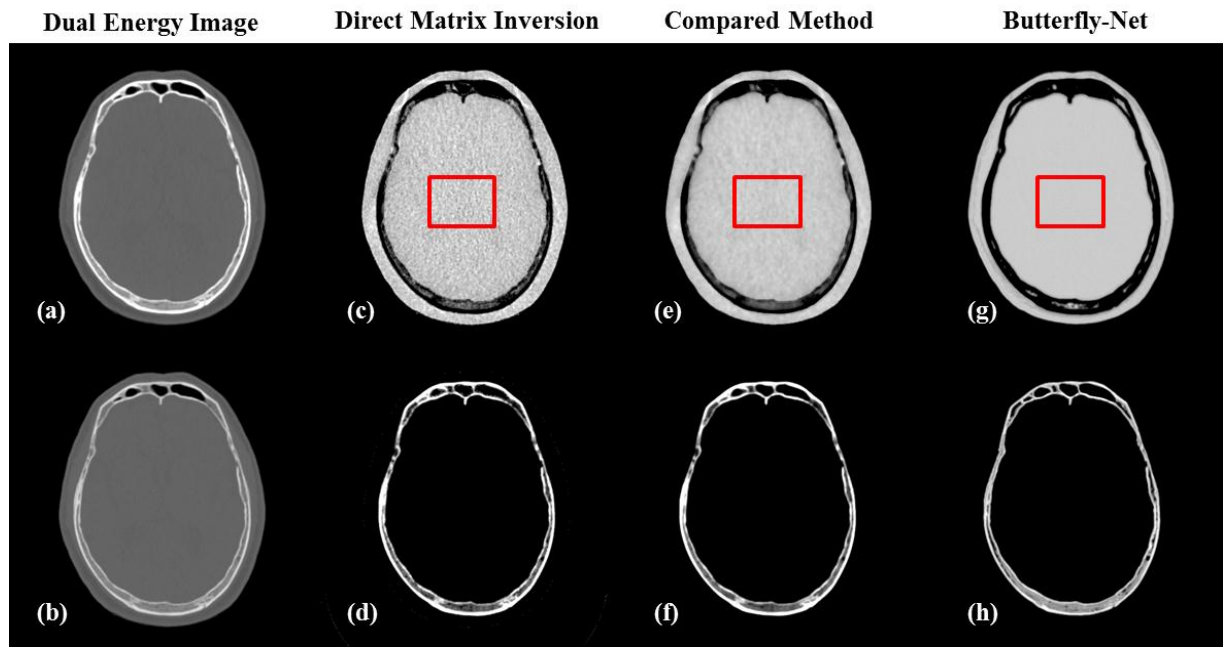

Figure 9 Bone and tissue decomposition results of group 1. (a) and (b) are the low- and high-energy images at tube voltages of 80 and 140 KVp with a display window of [0 0.05] mm$^{-1}$, respectively. (c) and (d) represent the decomposition results of direct matrix inversion. (e) and (f) represent the decomposition results of the compared method. (g) and (h) represent the results of Butterfly-Net. The display window of (c)–(h) is [0.2 1.2]. Red rectangle represents the selected ROI for quantitative analysis.

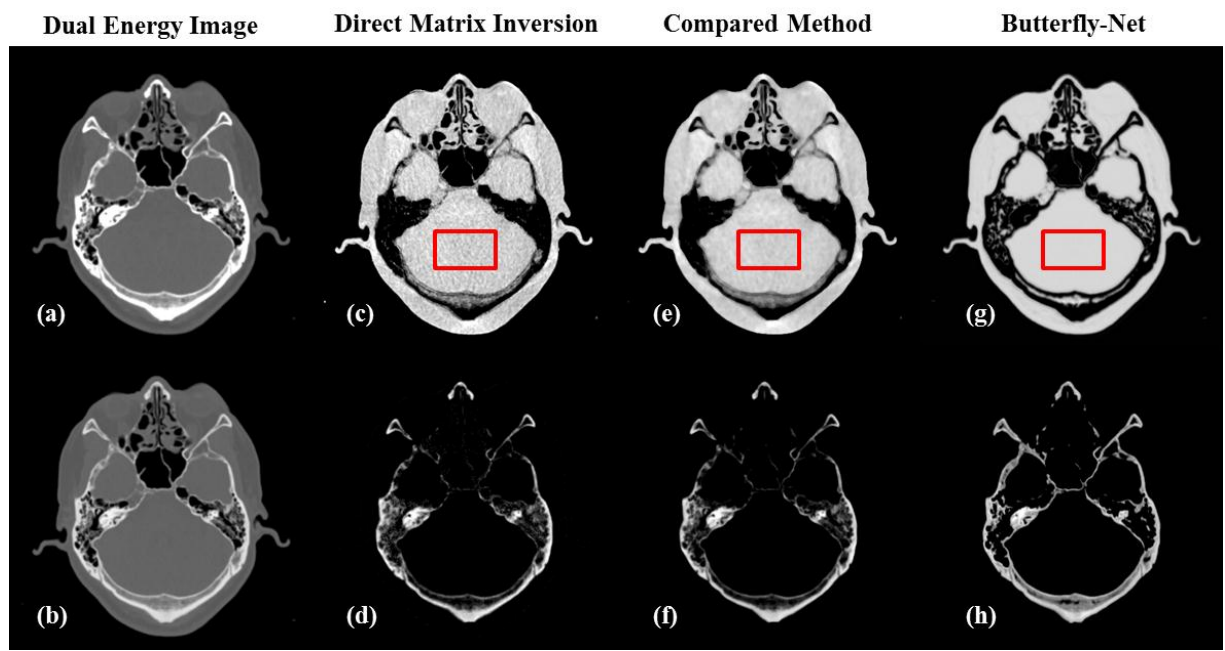

Figure 10 Bone and tissue decomposition results of group 2. (a) and (b) are the low and high-energy images at tube voltages of 80 and 140 KVp with display window [0 0.05] mm$^{-1}$, respectively. (c) and (d) represent the decomposition results of direct matrix inversion. (e) and (f) represent the decomposition results of compared method. (g) and (h) represent the results of Butterfly-Net. The display window of (c)–(h) is [0.2 1.2]. Red rectangle represents the selected ROI for quantitative analysis.

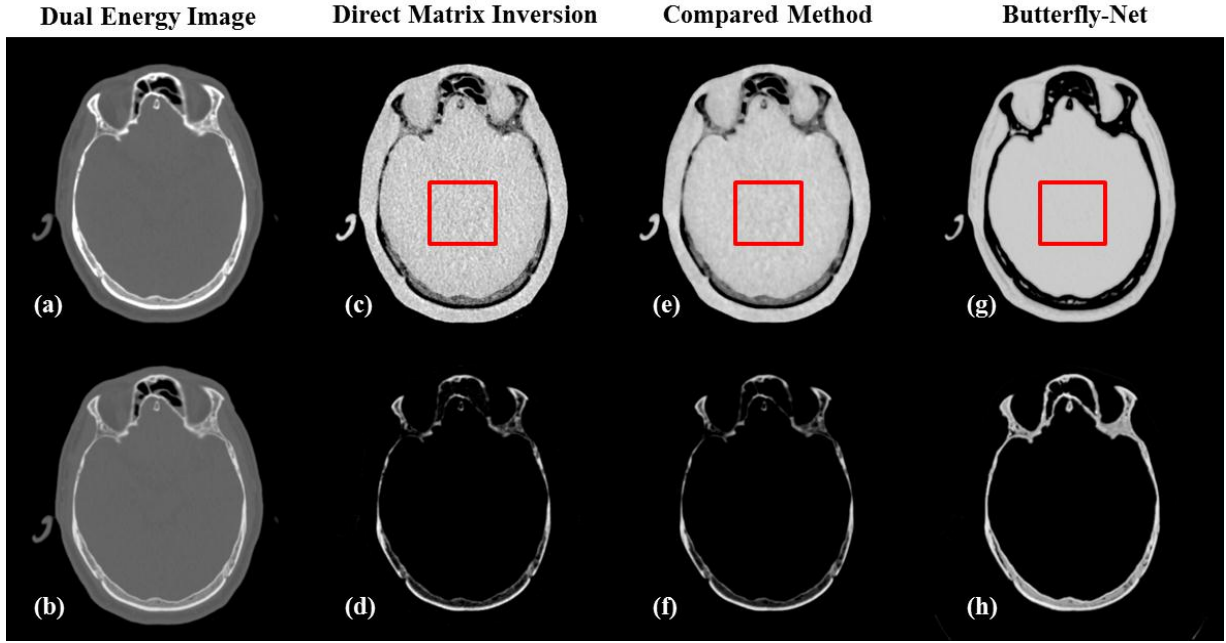

Figure 11 Bone and tissue decomposition results of group 3. (a) and (b) are the low and high-energy images at tube voltages of 80 and 140 KVp with display window [0 0.05] mm$^{-1}$, respectively. (c) and (d) represents the decomposition results of direct matrix inversion. (e) and (f) represent the decomposition results of compared method. (g) and (h) represent the results of Butterfly-Net. The display window of (c)–(h) is [0.2 1.2]. Red rectangle represents the selected ROI for quantitative analysis.

Table 1 Quantitative Results of Decomposition.

| Algorithm<br>Patient Image | Direct Inversion | Compared Method | Butterfly-Net |
|---|---|---|---|
| Group 1 | 0.9968 ±0.0593 | 0.9963 ±0.0214 | **0.9963 ±0.0040** |
| Group 2 | 1.0119 ±0.0492 | 1.0119 ±0.0239 | **1.0119 ±0.0034** |
| Group 3 | 0.9939 ±0.0515 | 0.9940 ±0.0228 | **0.9940 ±0.0045** |

## V. CONCLUSION AND DISCUSSION

DECT plays an important role in advanced CT applications due to its potential for differentiating materials. However, significant noise boosting in the signal decomposition process severely degrades the quality of decomposed material images. This paper proposed a Butterfly-Net derived from the image domain decomposition model of DECT to decompose basis materials from dual-energy images. The visualization evaluation indicated the roles of network components in material decomposition and clarified how the integrated network performs material decomposition. The Butterfly-Net exhibited excellent performance in noise suppression and exhibited increased quality on image domain material decomposition for DECT.

In this work, the decomposition model is transformed into the Butterfly-Net by exploring the corresponding relationship between mapping functions of DECT and network components. This new network establishes crossover architecture between two material generation pathways and it actually plays a key role in the network. Neurons of crossover architecture show different sensitivities to different basis materials, enabling the integrated network to learn and recognize different material information from high- and low-energy images. Without the crossover architecture, information exchange cannot be performed between two material generation pathways, and material decomposition is also impossible. Regardless of the network designed for material decomposition, information exchange between high- and low-energy images is unavoidable and is determined by the physical mechanism of data model. We may continually design some networks with single input and single output without containing crossover architecture. When we further analyze the working mechanism of network, some assembly neurons aiming to exchange information between high- and low-energy images will be presented during the decomposition process. In this case, we can also consider these hidden-state neurons as a crossover architecture, which may not be intuitive as the architecture of our network. Therefore, we can conclude that the hidden/intuition crossover network is necessary for DECT regardless whether the network is single-entry single-output or multi-entry multi-output.

The proposed network is actually not a strict expression of the image domain decomposition model. Instead, the data model is the basis of the new network and the mechanism of data model provides a guideline for the network design. The generalization and containments of neural network enable the Butterfly-Net superior performance in image domain material decomposition for DECT. Our network is the first step of applying CNN to image domain material decomposition for DECT. The tube voltages are fixed at 80 and 140 KVp in this paper and we need to establish other datasets and retrain the network if we prefer to perform material decomposition at different energies. Fortunately, for a specific DECT system, fixed tube voltages and datasets are desirable because its energy spectra and scanning parameters will not change frequently. The network structure presented in this paper is also flexible and its performance can be further improved by exploring additional hidden stages of the network. However, a key problem that must be considered is that a deep network requires large memory and long time for training. In addition, we may find that the decomposition results of our network are not completely consistent with the compared method for the image pixels in the adjacent area of bone and tissue. This is because the datasets established in this paper is not extensive enough for network training. We can continually expand training datasets to million samples or introduce regularization-based approach into the network to raise the ability of the network.

# ACKNOWLEDGEMENTS

This work was supported by the National Natural Science Foundation of China (No. 61601518) and national key research and development plan of China under grant (2017YFB1002502, 2016YFC0104507). It is also partially supported by Zhejiang Provincial Natural Science Foundation of China (Grant No. LR16F010001, LY16H180001), National High-tech R&D Program for Young Scientists by the Ministry of Science and Technology of China (Grant No. 2015AA020917), Zhejiang Province 151 Talents Program.

# REFERENCE

1. W. A. Kalender, W. H. Perman, J. R. Vetter, and E. Klotz, "Evaluation of a prototype dual-energy computed tomographic apparatus. I. Phantom studies," Med. Phys. **13**(3), 334–339(1986).

2. T. Johnson, B. Krau, M. Sedlmair, M. Grasruck, H. Bruder, D. Morhard, C. Fink, S. Weckbach, M. Lenhard, B. Schmidt, T. Flohr, M. F. Reiser, and C. R. Becker, "Material differentiation by dual energy CT: initial experience," European J. Radiology **17**(6), 1510–1517(2007).

3. S. Singh and M. Singh, "Explosives detection systems (EDS) for aviation security," Signal Process. **83**(1), 31–55(2003).

4. Z. Ying, R. Naidu, and C. R. Crawford, "Dual energy computed tomography for explosive detection," J. X-Ray Sci. Tech. **14**(4), 235–256(2006).

5. P. Engler and W. Friedman, "Review of dual-energy computed tomography techniques," Mater. Eval. **48**(5), 623–629(1990).

6. L. Y. C. McCollough, S. Leng and J. Fletcher, "Scikit-image: Dual- and multi-energy ct: principles, technical approaches, and clinical applications," Radiology **276**(3), 637–653(2015).

7. R. Zhang, J. B. Thibault, C. Bouman, K. Sauer, and J. Hsieh, "Model-based iterative reconstruction for dual-energy x-ray CT using a joint quadratic likelihood model" IEEE Trans. Med. Imaging **33**(1), 117-134(2014).

8. R. E. Alvarez and A. Macovski, "Energy-selective reconstructions in x-ray computerized tomography," Med. Phys. **21**(5), 733–744(1976).

9.  L. A. Lehmann, R. E. Alvarez, A. Macovski, W. R. Brody, N. J. Pelc, S. J. Riederer, and A. L. Hall, "Generalized image combinations in dual KVP digital radiography," Med. Phys. **8**(5), 659–667(1981).

10. B. R. Foygel, E. Y. Sidky, S. T. Gilat, and X. Pan, "An algorithm for constrained one-step inversion of spectral CT data," Phy. Med. Biol. **61**(10), 3784–3818(2015).

11. Y. Long, J.A. Fessler, "Multi-material decomposition using statistical image reconstruction for spectral CT," IEEE Trans. Med. Imaging **33**(8), 1614–1626(2014).

12. B. Chen, Z. Zhang, E. Y. Sidky, D. Xia, and X. Pan, "Image reconstruction and scan configurations enabled by optimization based algorithms in multispectral CT," Phy. Med. Biol. **62**(22), 8763–8793(2017).

13. Y. Xue, R. Ruan, X. Hu, Y. Kuang, J. Wang, Y. Long, and T. Niu, "Statistical image-domain multi-material decomposition for dual-energy CT," Med. Phys. **44**(3), 886–901(2017).

14. D. S. Rigie and P. J. LaRiviere, "Joint reconstruction of multi-channel, spectral CT data via constrained total nuclear variation minimization," Phys. Med. Biol. **60**(5), 1741–1762(2015).

15. E.Y. Sidky, Y. Zou, X. Pan, "Impact of polychromatic x-ray sources on helical, cone-beam computed tomography and dual-energy methods," Phy. Med. Biol. **49**(11), 2293–2303(2004).

16. M. Daniele, T. B. Daniel, M. Achille, C. N. Rendon, "State of the art: Dual-energy CT of the abdomen," Radiology **271**, 327–342(2014).

17. J. Harms, T. Wang, M. Petrongolo, T. Niu, L. Zhu, "Noise suppression for dual-energy CT

via penalized weighted least-square optimization with similarity-based regularization," Med. Phy. **43**(5), 2676–2686(2016).

18. P. Sukovic and N. H. Clinthorne, "Penalized weighted least-squares image reconstruction for dual energy x-ray transmission tomography," IEEE Trans. Med. Imaging **19**(11), 1075–1081(2000).

19. C. Maass, M. Baer, and M. Kachelriess, "Image-based dual energy CT using optimized precorrection functions: A practical new approach of material decomposition in image domain," Med. Phys. **36**(8), 3818–3829(2009).

20. F. Kelcz, P. M. Joseph, and S. K. Hilal, "Noise considerations in dual energy CT scanning," Med. Phys. **6**(5), 418–425 (1979).

21. A. Macovski, D. G. Nishimura, A. Doost-Hoseini, W. R. Brody, "Measurement-dependent filtering: A novel approach to improved SNR," IEEE Trans. Med. Imaging **2**(3), 122–127 (1983).

22. P. C. Johns and M. J. Yaffe, "Theoretical optimization of dual-energy x-ray imaging with application to mammography," Med. Phys. **12**(3), 289–296 (1985).

23. R. A. Rutherford, B. R. Pullan, and I. Isherwood, "X-ray energies for effective atomic number determination," Neuroradiology **11**(1), 23–28 (1976).

24. R. J. Warp and J. T. Dobbins, "Quantitative evaluation of noise reduction strategies in dual-energy imaging," Med. Phys. **30**(2), 190–198 (2003).

25. A. Hinshaw and J. T. Dobbins, "Recent progress in noise reduction and scatter correction in dual-energy imaging," Proc. SPIE **2432**, 134–142 (1995).

26. T. Niu, X. Dong, M. Petrongolo, L. Zhu, "Iterative image-domain decomposition for

dual-energy CT," Med. Phy. **41**(4), 041901(2014).

27. G. Wang, "A perspective on deep imaging," IEEE Access **4**(99), 8914-8924(2017).

28. D. Wu, K. Kim, G. E. Fakhri, and Q. Li, "Iterative low-dose CT reconstruction with priors trained by artificial neural network," IEEE Trans. Med. Imaging **36**(12), 2479–2486(2017).

29. H. Chen, Y. Zhang, M. K. Kalra, F. Lin, Y. Chen, P. Liao, J. Zhou, and G. Wang, "Low-dose CT with a residual encoder-decoder convolutional neural network," IEEE Trans. Med. Imaging **36**(12), 2524–2535(2017).

30. J. M. Wolterink, T. Leiner, M. A. Viergever, and I. Isgum, "Generative adversarial networks for noise reduction in low-dose CT", IEEE Trans. Med. Imaging **36**(12), 2536–2545(2017).

31. H. Chen, Y. Zhang, W. Zhang, P. Liao, K. Li, J. Zhou, and G. Wang, "Low-dose CT via convolutional neural network," Biomed. Opt. Express **8**(2): 679–694(2017).

32. E. Kang, J. Min, and J. C. Ye, "A deep convolutional neural network using directional wavelets for low-dose x-ray ct reconstruction," Med. Phys. **44**(10), e360–e375(2017).

33. E. Kang, J. Min, and J. C. Ye, "Wavelet domain residual network (WavResNet) for low-dose x-ray CT reconstruction," arXiv: 1703.01383v1, 2017.

34. K. H. Jin, M. T. McCann, E. Froustey, and M. Unser, "Deep convolutional neural network for inverse problems in imaging," IEEE Trans. Image Process. **26**(9), 4509–4522(2017).

35. J. Adler and O. Oktem, "Learned primal-dual reconstruction," arXiv:1707.06474v1, 2017.

36. D. Wu, K. Kim, B. Dong, and Q. Li, "End-to-end abnormality detection in medical imaging," arXiv: 1711.02074, 2017.

37. Y. Lu, A. Zhong, Q. Li, and B. Dong, "Beyond finite layer neural networks: Bridging deep architectures and numerical differential equations," arXiv: 1710.10121, 2017.

38. Y. Liu, "Channel codes," Zhengzhou, Henan Science and Technology Press, 2006.

39. T. P. Szczykutowicz and G. H. Chen, "Dual energy CT using slow kVp switching acquisition and prior image constrained compressed sensing," Phys. Med. Biol. **55**(21), 6411–6429 (2010).

40. K. Gregor and Y. LeCun, "Learning Fast Approximations of Sparse Coding," *International Conference on Machine Learning*, (Haifa, Israel, 2010), pp. 399–406.

41. B. Mark and S. Philip, "Onsager-corrected deep learning for sparse linear inverse problems," arXiv:1607.05966v1, 2016.

42. Z. Wang, Y. Yang, S. Chang, Q. Ling, and T. S. Huang, "Learning a deep $l_\infty$ encoder for hashing," arXiv:1604.01475v1, 2016.

43. K. He, X. Zhang, S. Ren, and J. Sun, "Deep Residual Learning for Image Recognition," arXiv:1512.03385v1, 2015.

44. H. Zhang, L. Li, K. Qiao, L. Wang, B. Yan, L. Li, and G. Hu, "Image predication for limited-angle tomography via deep learning with convolutional neural network," arXiv:1607.08707, 2016.

45. H. Zhang, L. Wang, Y. Duan, L. Li, G. Hu, and B. Yan, "Euler's Elastica Strategy for Limited-angle Computed Tomography Image Reconstruction," IEEE Trans. Nucl. Sci. **64**(8), 2395-2405(2017).

46. K. Simonyan, A. Vedaldi, and A. Zisserman, "Deep inside convolutional networks: Visualising image classification models and saliency maps," *International Conference on*

*Learning Representations Workshop*, (Banff, Canada, 2014).

47. W. Samek, A. Binder, G. Montavon, S. Bach, and K. R. Müler, "Evaluating the visualization of what a Deep Neural Network has learned," IEEE Trans. Neur. Net. Lear. **28**(11), 2660–2673(2016).